\documentclass[12pt]{article}

\usepackage{graphicx}
\usepackage{mathptmx}           % use Times fonts if available on your TeX system
\usepackage[latin1]{inputenc} % ansinew ou unicode in Linux
\usepackage{dsfont}
\usepackage{amsmath, amsthm, amssymb}
\usepackage{clrscode}             % Cormen & al. pseudocode package
\usepackage{url} 
\usepackage{color}

\newtheorem{theorem}{Theorem}
\newtheorem{lemma}[theorem]{Lemma}

\newtheorem*{Loop Invariant}{Loop Invariant}{\bf}{\it}

\begin{document}
\sloppy

\title{\bf An Optimized Divide-and-Conquer Algorithm for the Closest-Pair Problem in the Planar Case}

\author{   {\bf José C. Pereira}\\
            \small DEEI-FCT\\
            \small Universidade do Algarve\\
            \small Campus de Gambelas\\
            \small 8005-139 Faro, Portugal\\
            \small unidadeimaginaria@gmail.com
\and
		 {\bf Fernando G. Lobo}\\
            \small CENSE and DEEI-FCT\\
            \small Universidade do Algarve\\
            \small Campus de Gambelas\\
            \small 8005-139 Faro, Portugal\\
            \small fernando.lobo@gmail.com         
}
\date{}
\maketitle

\begin{abstract}
We present an engineered version of the \emph{divide-and-conquer} algorithm for finding the closest pair of points, within a given set of points in the XY-plane. For this version of the algorithm we show that only two pairwise comparisons are required in the \emph{combine} step, for each point that lies in the 2$\delta$-wide vertical slab. The correctness of the algorithm is shown for all Minkowski distances with $p\geqslant 1$. We also show empirically that, although the time complexity of the algorithm is still $O(n~lg~n)$, the reduction in the total number of comparisons leads to a significant reduction in the total execution time, for inputs with size sufficiently large. 
\end{abstract}

\section{Introduction}\label{sec:intro}
The Closest-Pair problem is considered an ``easy'' Closest-Point problem, in the sense that there are a number of other geometric problems (e.g. nearest neighbors and minimal spanning trees) that find the closest pair as part of their solution~\cite[p. 226]{1}. This problem and its generalizations arise in areas such as statistics, pattern recognition and molecular biology. 

At present time, many algorithms are known for solving the Closest-Pair problem in any dimension $k\geqslant2$, with optimal time complexity (see~\cite{2} for an overview of Closest-Point problem algorithms and generalizations). The Closest-Pair is also one of the first non-trivial computational problems that was solved efficiently using the \emph{divide-and-conquer} strategy and it became since a classical, textbook example for this technique. 

In this paper we consider only algorithms for the Closest-Pair problem that can be implemented in the \emph{algebraic computation tree model}. For this model any algorithm has time complexity $\Omega(n~lg~n)$. With more powerful machine models, where \emph{randomization}, the \emph{floor function}, and \emph{indirect addressing} are available, faster algorithms can be designed~\cite{2}.

\subsection*{Historical Background}\label{subsec:history}
An algorithm with optimal time complexity $O(n~lg~n)$ for solving the Closest-Pair problem in the planar case appeared for the first time in 1975, in a computational geometry classic paper by Ian Shamos~\cite{3}. This algorithm was based on the Voronoi polygons.

The first optimal algorithm for solving the Closest-Pair problem in any dimension $k\geqslant2$ is due to Jon Bentley and Ian Shamos~\cite{4}.  Using a \emph{divide-and-conquer} approach to initially solve the problem in the plane\footnote{According to Jon Bentley~\cite{1}, Shamos attributes the discovery of this procedure to H.R. Strong.}, those authors were able to generalize the planar process to higher dimensions by exploring a sparsity condition induced over the set of points in the $k$-plane.

For the planar case, the original procedure and other versions  of the \emph{divide-and-conquer} algorithm  usually compute at least seven pairwise comparisons for each point in the central slab, within the combine step (see \cite{4}, \cite{5}, and \cite{6}, for instance). 

In 1998, Zhou, Xiong, and Zhu\footnote{The article in question was published in chinese. See \cite{7} and \cite{8} for some explicit references.} presented an improved version of the planar procedure, where at most four pairwise comparisons need to be considered in the combine step, for each point lying on the left side (alternatively, on the right side) of the central slab. In the same article, Zhou \textit{\textit{et al}.} introduced the ``complexity of computing distances'', which measures ``the number of Euclidean distances to compute by a closest-pair algorithm''~\cite{7}. The core idea behind this definition is that, since the Euclidean distance is usually more expensive than other basic operations, it may be possible to achieve significant efficiency improvements by reducing this complexity measure.

More recently, Ge, Wang, and Zhu used some sophisticated geometric arguments to show that it is always possible to discard one of the four pairwise comparisons in the combine step, thus reducing significantly the complexity of computing distances, and presented their enhanced version of the Closest-Pair algorithm, accordingly~\cite{8}. 

In 2007, Jiang and Gillespie presented another version of the Closest-Pair \emph{divide-and-conquer} algorithm which reduced the complexity of computing distances by a logarithmic factor. However, after performing some algorithmic experimentation, the authors found that, albeit this reduction, the new algorithm was ``the slowest among the four algorithms''~\cite{7} that were included in the comparative study. The experimental results also showed that the fastest among the four algorithms was in fact a procedure named Basic-2, where two pairwise comparisons are required in the combine step, for each point that lies in the central slab and, therefore, has a relative high complexity of computing distances. The authors conclude that the simpler design in the combine step, and a consequent correct imbalance in trading expensive operations with cheaper ones are the main factors for explaining the success of the Basic-2 algorithm.

\begin{figure*}[t]
\centering
\fbox{
\includegraphics[width=15cm]{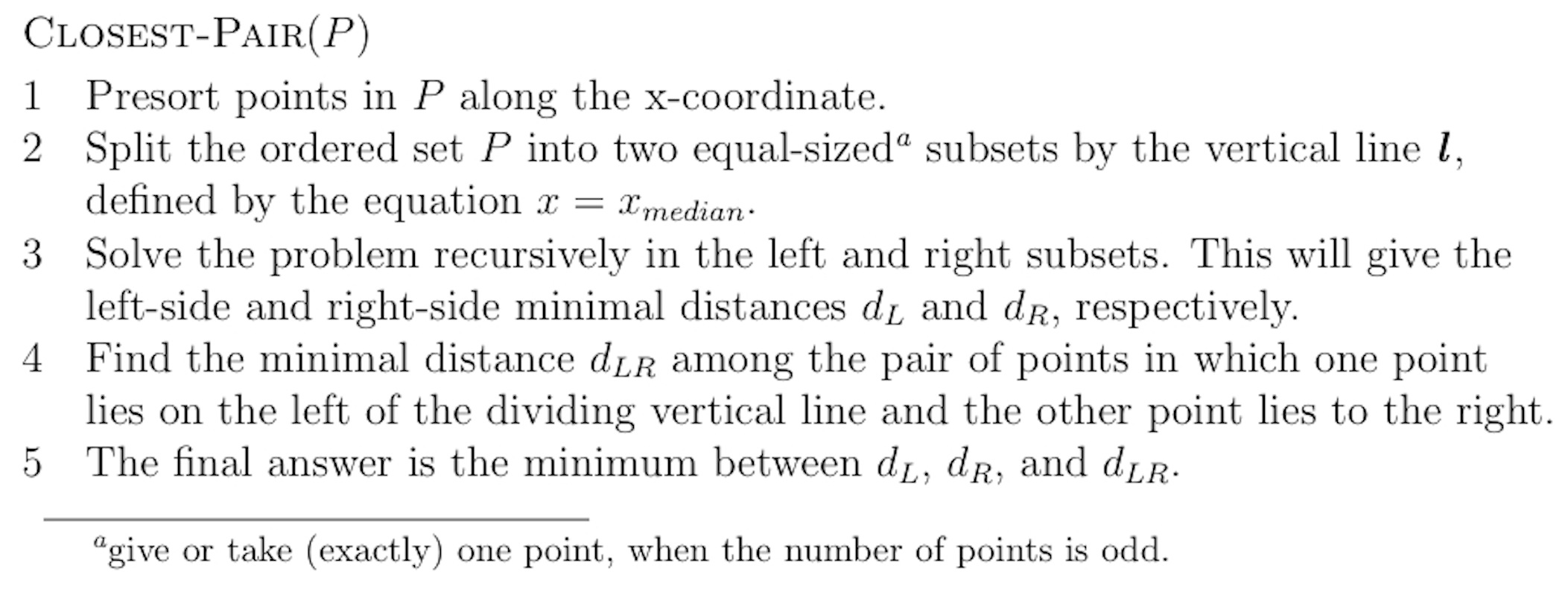}
}
\caption{Pseudocode for the divide-and-conquer Closest-Pair algorithm, first presented by Bentley and Shamos~\cite{4}.}
\label{fig:pscode1}
\end{figure*}

In this paper we present a detailed version of the Basic-2 algorithm. We show that, for this algorithm, only two pairwise comparisons are required in the combine step, for each point that lies in the central slab. This result and the subsequent correctness of the Basic-2 algorithm is shown for all Minkowski distances with $p\geqslant 1$.

In fairness to all parts involved, we must state that the main results presented in this paper, including the design of the Basic-2 algorithm, were obtained in a completely independent fashion from any previous work by other authors. It was only during the process of putting our ideas into writing that we came across with the articles by Zhou \textit{et al}, Qi Ge \textit{et al}~\cite{8}, and Jiang and Gillespie~\cite{7} which, obviously, take precedence and deserve due credit. Nonetheless, this paper gives a significant contribution to the Closest-Pair problem,  by establishing the correctness of the Basic-2 algorithm for all Minkowski distances with $p\geqslant 1$, in particular for the Euclidean distance.

The rest of the paper is organized as follows. In Section \ref{sec:closest-pair}, we review the classic Closest-Pair algorithm as presented by Bentley and Shamos~\cite{4}. In Section~\ref{sec:basic2}, we present our detailed version of the Basic-2 algorithm and give the correspondent proof of correctness. In Section~\ref{sec:empiric}, we present a comparative empirical study between the classic Closest-Pair and the Basic-2 algorithm, and discuss the experimental results obtained with distinct Minkowski distances.

\section{The divide-and-conquer algorithm in the plane}\label{sec:closest-pair}

The following algorithm for solving the planar version of the Closest-Pair problem was first presented by Bentley and Shamos~\cite{4}.

Let $P$ be a set of $n\geq2$ points in the XY-plane. The closest pair in $P$ can be found in $O(n~lg~n)$ time using the \emph{divide-and-conquer} algorithm shown in Figure~\ref{fig:pscode1}.

Since we are splitting a set of $n$ points in two sets of $n/2$ points each, the recurrence relation describing the running time of the Closest-Pair algorithm is $T(n) = 2T(n/2) + f(n)$, where $f(n)$ is the running time for finding the distance $d_{LR}$ in step 4. 

At first sight it seems that something of the order of $n^2/4$ distance comparisons will be required to compute $d_{LR}$. However, Bentley and Shamos~\cite{4} noted that the knowledge of  both distances $d_L$ and $d_R$ induces a sparsity condition over the set $P$.

Let $\delta = min(d_{L},d_{R})$ and consider the vertical slab of width $2\delta$ centered at line \textbf{\textit{l}}. If there is any pair in $P$ closer than $\delta$, both points of the pair must lie on opposite sides within the slab. Also, because the minimum separation distance of points on either side of \textbf{\textit{l}} is $\delta$,  any square region of the slab, with side $2\delta$, ``can contain at most a constant number \textbf{c} of points''~\cite{4}, depending on the used metric\footnote{In the original article, Bentley and Shamos~\cite{4} used the Minkowski distance $d_\infty$ and obtained the value $\textbf{c}=12$.}.

As a consequence of this sparsity condition, if the points in $P$ are presorted by y-coordinate, the computation of $d_{LR}$ can be done in linear time. Therefore, we obtain the recurrence relation $T(n) = 2T(n/2) + O(n)$, giving an $O(n~lg~n)$ asymptotically optimal algorithm.

\begin{figure}
\centering
\includegraphics[width=7.5cm]{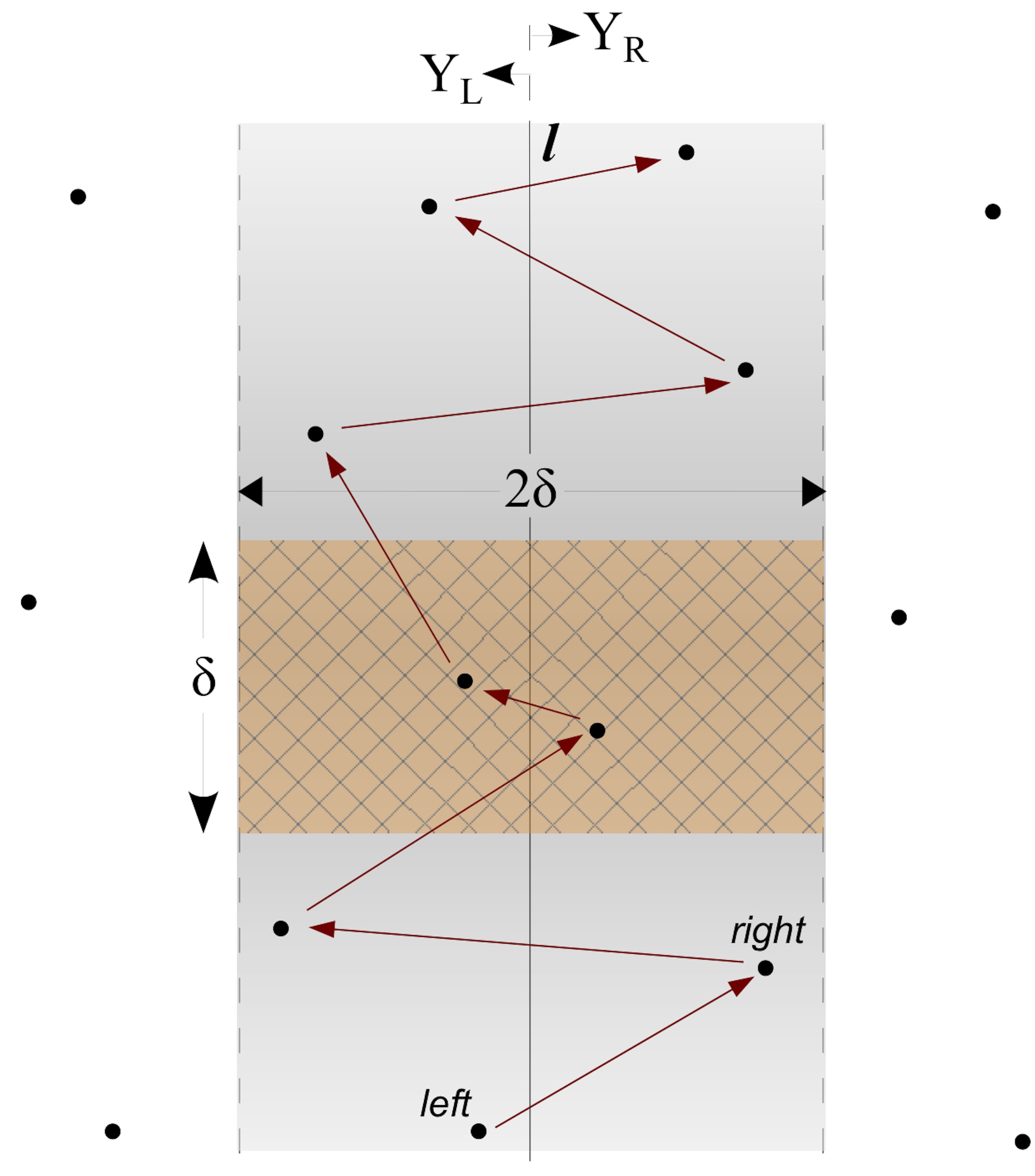}
\caption{``Hopscotching'' $\Delta$-points in ascending order. For each point visited on either side, the Basic-2.S4 algorithm computes the distance for the two closer, but not lower, points on the opposite side.	}
\label{fig:hopScotch}
\end{figure}

\section{The Basic-2 algorithm}\label{sec:basic2}

\begin{figure*}[t]
\centering
\fbox{
\includegraphics[width=12cm]{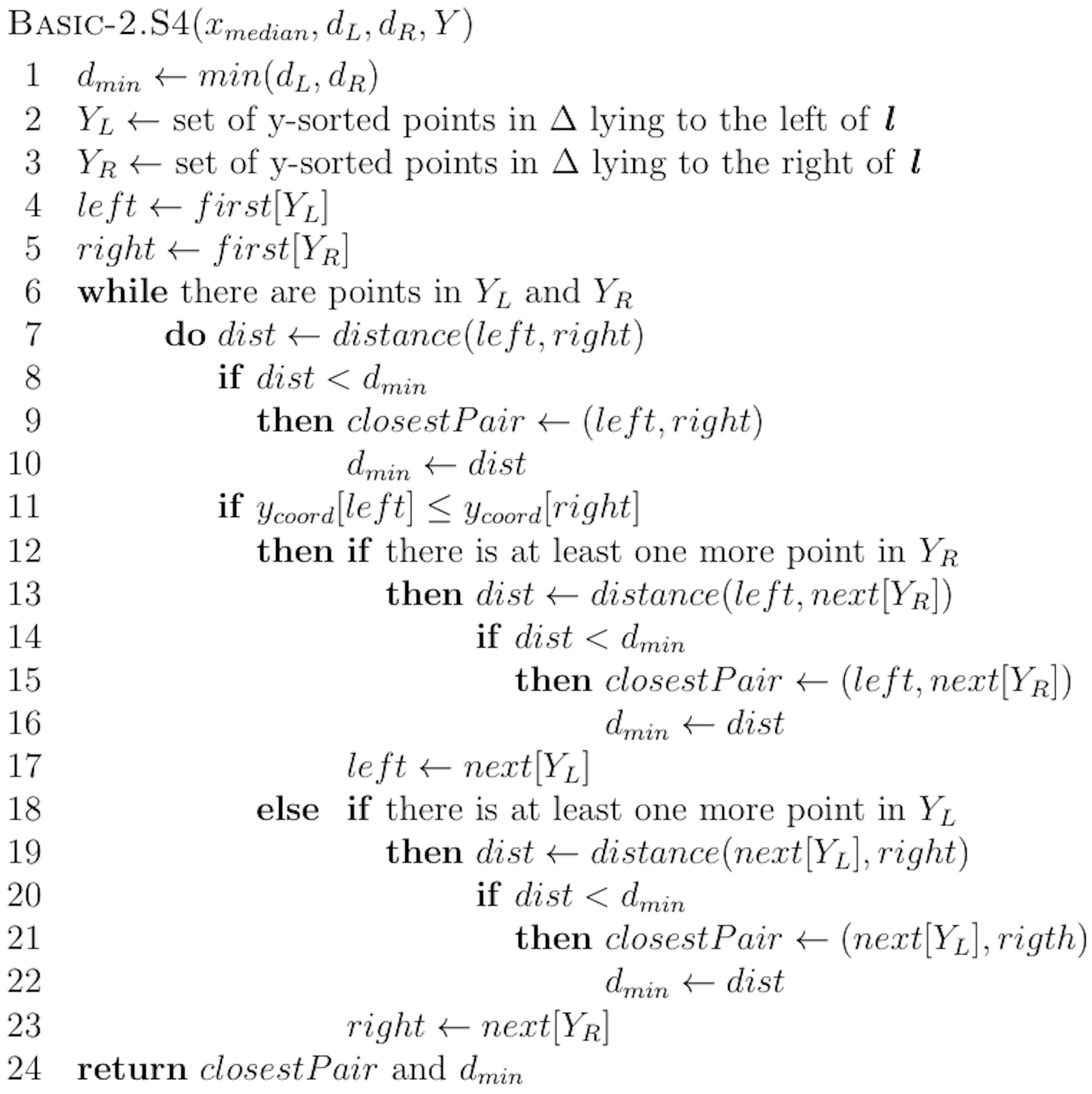}
}
\caption{Pseudocode for step 4 of the Basic-2 algorithm.Note that \textit{next}[] is a reference to the point immediately above point {\it left (right)}, in the ascending y-sorted set $Y_L$ ($Y_R$). This reference does not change those sets.}
\label{fig:pscode2}
\end{figure*}

In this Section we discuss a detailed version of the Basic-2 algorithm, which was first presented by Jiang and Gillespie~\cite{7}. 

The Basic-2 algorithm is an optimized version of the Bentley and Shamos procedure for the planar case discussed in Section~\ref{sec:closest-pair}. In fact, the Basic-2 algorithm is the same as the Closest-Pair algorithm (see Figure~\ref{fig:pscode1}), with the sole difference that the computation of the distance $d_{LR}$ in step 4 now requires only two pairwise comparisons per point to find the closest pair within the central slab. The pseudocode for computing the $d_{LR}$ distance in the Basic-2 algorithm is shown in Figure~\ref{fig:pscode2}.

The time complexity of the Basic-2.S4 algorithm is obviously $O(n)$, since it traverses once the arrays $Y_L$ and $Y_R$ in a ``hopscotch'' manner (see Figure~\ref{fig:hopScotch}), and it takes constant time on each iteration.

Since we are only interested in performing step 4 of the Closest-Pair algorithm, in the following we assume that $x_{median}$, $d_{L}$, and $d_{R}$ are already computed. We also assume that the array $Y$ contains a y-sorted partition of all points in $P$, i.e., the first and second halves of $Y$ contain the points that are to the left and to the right of \textbf{\textit{l}}, respectively, and both halves are sorted along the y-coordinate\footnote{The structure of the array $Y$ may seem difficult to obtain, and therefore, be an extra source of complexity in the overall algorithm. However, this is not the case because the structure arises as a natural consequence from the need to maintain the y-presorting throughout the recursive calls in $O(n)$ time.}.

We denote the vertical slab centered at line $x = x_{median}$ of width $2\delta$ by the symbol $\Delta$ and the central line by \textbf{\textit{l}}. 

Before we show the correctness of the algorithm we first prove the following\\

\begin{lemma}\label{lemma1}
Let $d_p:\mathds{R}^2 \times \mathds{R}^2 \rightarrow \mathds{R}$ denote the Minkowski p-distance, $1
\leqslant p \leqslant \infty$. Let $P_0=(x_0,y_0) \in Y_L (respectively, Y_R)$ be an arbitrary point lying in the central slab $\Delta$, and let $Y_0\subseteq Y_R (respectively, Y_L)$ be the array of points that lie opposite and above $P_0$, sorted along the y-coordinate. The closest point to $P_0$, in $Y_0$, with respect to $d_p$, is either the first or the second element of $Y_0$.
\end{lemma}

\

\begin{proof}   
We first give proof for the Minkowski distance for $p=1$, defined by $d_1(A,B)=|x_A-x_B|+|y_A-y_B|$. We assume, without loss of generality, that $P_0=(0,0) \in Y_L$ and, as a consequence, that $Y_0 \subseteq Y_R$. Let $A=(a,b), ~0\leqslant a \leqslant \delta, ~b \geqslant 0$ be the first point in $Y_0$. We note that, because $Y_0$ is sorted along the y-coordinate, it is sufficient to consider the case where $b=0$, since all others cases with $b>0$ can be obtained by making an upper translation. This translation does not disrupts the relative positions of the elements in $Y_0$ and therefore, all arguments presented for $b=0$ will remain valid. So, let $A=(a,0)$ and let $P=(x,y), ~0\leqslant x\leqslant \delta, ~y\geqslant0$, be any other point in $Y_0$. We must consider three cases (ilustrated in Figures \ref{fig:lemma1a}, \ref{fig:lemma1b}, and \ref{fig:lemma1c}).

\begin{figure}[h!]
\centering
\includegraphics[scale=.4]{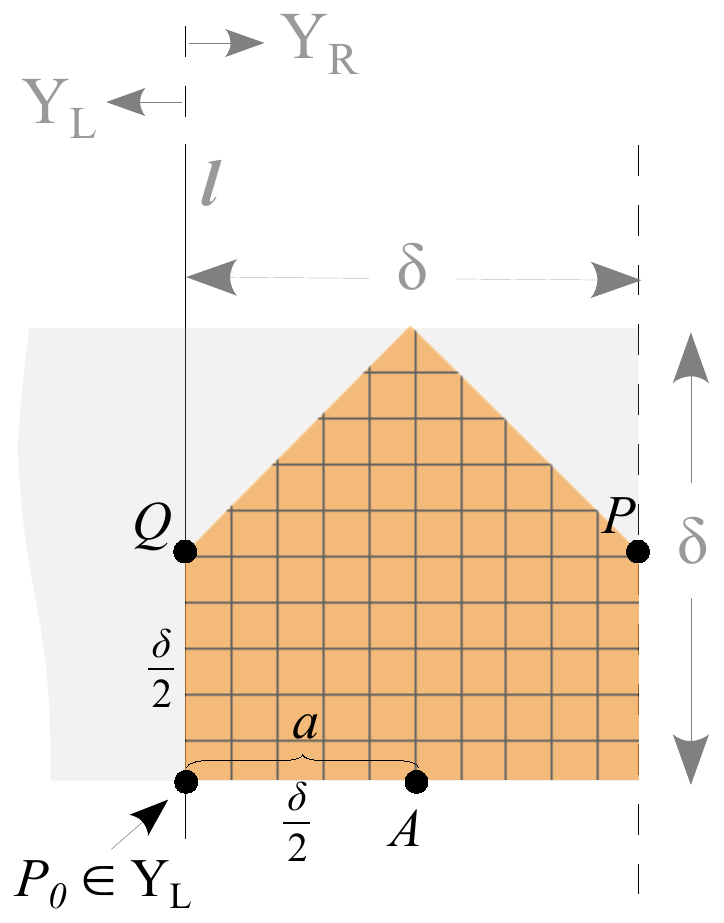}
\caption{Possible location for the first three  points, $A$, $P$, and $Q$ in Case 1. When $A$ is centered it is possible for both $A$ and $Q$ to be the closest points to $P_0$, in $Y_0$. However, this is a limit-case.}
\label{fig:lemma1a}
\end{figure}

\begin{description}
\item[Case 1:]  $a=\frac{\delta}{2}$. We find that
\begin{eqnarray*}
	&& 0\leqslant x\leqslant \delta \Leftrightarrow\\ 
     &\Leftrightarrow& 0-\frac{\delta}{2}\leqslant x-\frac{\delta}{2}\leqslant \delta-\frac{\delta}{2} \Leftrightarrow\\
    	&\Leftrightarrow& -\frac{\delta}{2}\leqslant x-\frac{\delta}{2}\leqslant \frac{\delta}{2} \Leftrightarrow\\
   	&\Leftrightarrow& |x-\frac{\delta}{2}|\leqslant \frac{\delta}{2}
\end{eqnarray*}

On the other hand, we have
\begin{eqnarray*}\label{P_theta_psi}
    && d_1(A,P) \geqslant \delta \Leftrightarrow\\ 
     &\Leftrightarrow& |x-\frac{\delta}{2}|+y \geqslant \delta \Leftrightarrow\\
    &\Leftrightarrow& y \geqslant \delta - |x-\frac{\delta}{2}| \geqslant  \delta-\frac{\delta}{2} = \frac{\delta}{2}
\end{eqnarray*}
Therefore,
$$d_1(P_0,P)=x+y \geqslant x + \frac{\delta}{2} \geqslant \frac{\delta}{2} = d_1(P_0,A)$$
which is to say that $A$ is a closest point to $P_0$, in $Y_0$.\\
We note that, if we take $y=\frac{\delta}{2}$, the first three points in $Y_0$ may have coordinates $A=(\frac{\delta}{2},0); ~P=(\delta,\frac{\delta}{2}); ~Q=(0,\frac{\delta}{2})$, respectively. This is the limit-case depicted in Figure \ref{fig:lemma1a}, where $A$ and $Q$ are both the closest points to $P_0$, in $Y_0$.\\

\begin{figure}[h!]
\centering
\includegraphics[scale=.4]{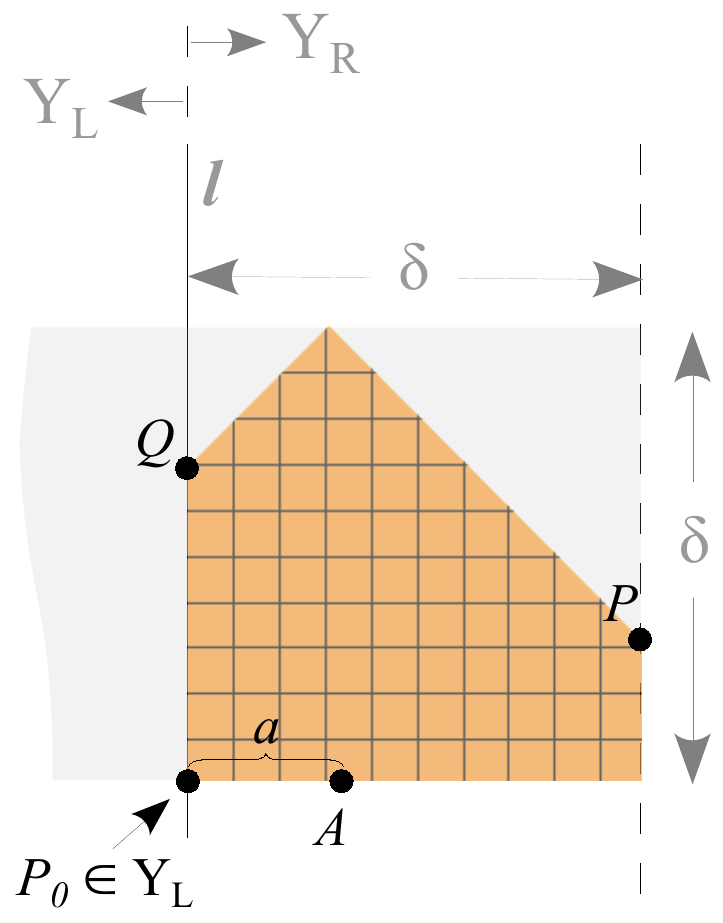}
\caption{Possible location for the first three  points, $A$, $P$, and $Q$ in Case 2. When $A$ is close to $P_0$ all other points in $Y_0$ are ``pushed'' away by the sparsity of $\Delta$.}
\label{fig:lemma1b}
\end{figure}

\

\

\item[Case 2:]   $0\leqslant a < \frac{\delta}{2}$. We consider two possibilities\\
\begin{enumerate}
\item[i)] Let $y > \frac{\delta}{2}$, then
$$d_1(P_0,P)=x+y > x+\frac{\delta}{2} \geqslant \frac{\delta}{2} > a = d_1(P_0,A).$$
\item[ii)] Let $y \leqslant \frac{\delta}{2}$, then
\begin{eqnarray*}
	&&d_1(A,P) \geqslant \delta \Leftrightarrow |x-a|+y \geqslant \delta \Leftrightarrow\\
	&\Leftrightarrow& |x-a| \geqslant \delta-y \geqslant \delta-\frac{\delta}{2}=\frac{\delta}{2} \Leftrightarrow\\
	&\Leftrightarrow& x-a \geqslant \frac{\delta}{2} ~\vee~ x-a \leqslant -\frac{\delta}{2} \Leftrightarrow\\
	&\Leftrightarrow& x \geqslant a+\frac{\delta}{2} ~\vee~ \underbrace{x \leqslant a-\frac{\delta}{2} < 0}_{Contradiction} \Rightarrow\\
	&\Rightarrow& x \geqslant a+\frac{\delta}{2}
\end{eqnarray*}
Therefore,
$$d_1(P_0,P)=x+y \geqslant a+\frac{\delta}{2}+y \geqslant \frac{\delta}{2} > a = d_1(P_0,A).$$
\end{enumerate}
Considering i) and ii) we conclude that A is the closest point to $P_0$, in $Y_0$.\\

\begin{figure}[h!]
\centering
\includegraphics[scale=.4]{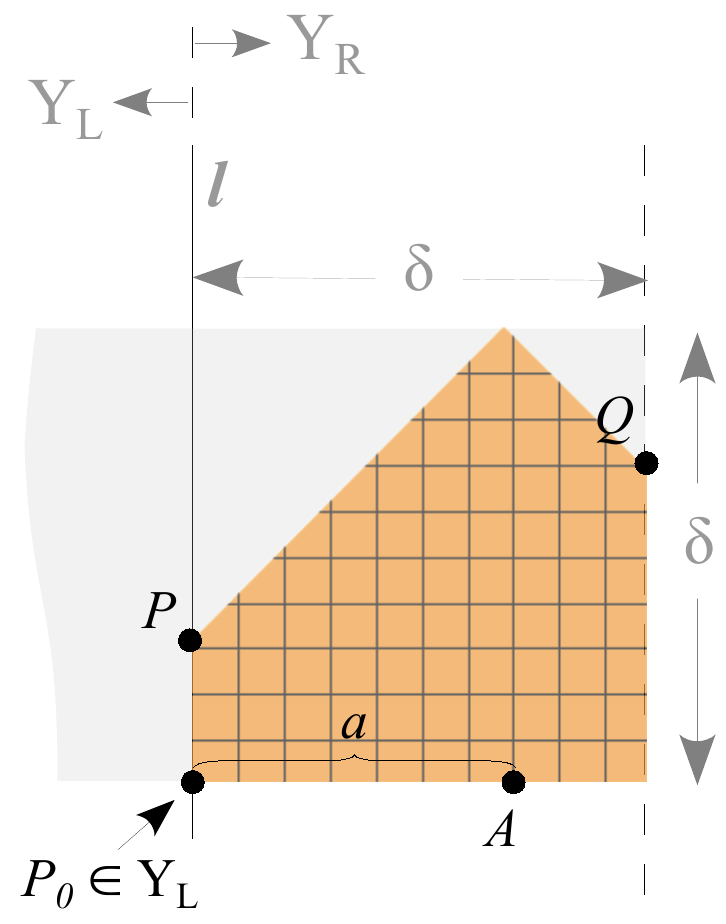}
\caption{Possible location for the first three  points, $A$, $P$, and $Q$ in Case 3. When $A$ lies further from $P_0$ it is possible for another point to be closer to $P_0$. However, this point is necessarily the next lowest point, which is to say, this point is the second element in $~Y_0$.}
\label{fig:lemma1c}
\end{figure}

\

\

\item[Case 3:]   $\frac{\delta}{2} < a \leqslant \delta$. We must consider two possibilities\\
\begin{enumerate}
\item[i)] Let $x \geqslant a$, then
$$d_1(P_0,P)=x+y \geqslant a+y \geqslant a = d_1(P_0,A).$$
and, as in the previous cases, $A$ is the closest point to $P_0$, in $Y_0$.\\
\item[ii)] Let $ x < a$, then
$$d_1(A,P) \geqslant \delta \Leftrightarrow a-x+y \geqslant \delta \Leftrightarrow y\geqslant \delta-a+x$$
This means that it is possible to have at least one point $P=(x,y) \in Y_0$ such that $d_1(P_0,P) = x+y < a=d_1(P_0,A)$, and $0 \leqslant x < a, ~\delta-a+x \leqslant y \leqslant a-x$. Let $Q=(x_1,y_1) \in Y_0$, and assume that $y_1 \leqslant y$ (i.e., Q precedes P in $Y_0$). We know that
\begin{eqnarray}\label{ineq1}
	\nonumber
	&&d_1(A,Q) \geqslant \delta \Leftrightarrow |x_1-a|+y_1 \geqslant \delta \Leftrightarrow\\
	&\Leftrightarrow& x_1 \geqslant a+ \delta -y_1 ~\vee~ x_1 \leqslant a- \delta +y_1
\end{eqnarray}
From the first inequality of (\ref{ineq1}) we find that
$$x_1 \geqslant a+ \delta -y_1 \geqslant a+ \delta -y > x +y + \delta -y = \delta + x \geqslant \delta$$
and this means that, in this case, for $Q$ to precede $P$ in $Y_0$, it must lie outside the $\Delta$ slab, which is a contradiction.

The second inequality of (\ref{ineq1}) holds only if we choose $\delta -a \leqslant y_1 \leqslant y$ to guarantee that $0 \leqslant x_1 \leqslant y_1 - (\delta -a)$. However, for this choice of the $Q$ coordinates, and from the sparsity of $\Delta$ we get\\
$$\setlength\arraycolsep{3pt}
 \begin{array}{lllll}
 	&d_1(P,Q) \geqslant \delta \Leftrightarrow |x_1 - x| + y ~- &y_1&\geqslant \delta \Leftrightarrow&\\
  	\Leftrightarrow& x_1 - x \geqslant \delta + y_1 -y &\vee& x_1 - x \leqslant -\delta + y - y_1 \Rightarrow&\\
	\Leftrightarrow& x_1 \geqslant x+ \delta + y_1 - (\delta - a + x) &\vee& x_1 \leqslant (x + y) -\delta - y_1 \Leftrightarrow&\\
	\Leftrightarrow& x_1 \geqslant y_1 + a &\vee& x_1 < a - \delta - y_1 \Leftrightarrow&\\
	\Leftrightarrow&  \underbrace{x_1 > y_1 + a - \delta}_{Contradiction} &\vee& x_1 < a - \delta \leqslant 0 &(2)
 \end{array} 
 $$
 
 The first inequality of (2) is a contradiction with our current working hypothesis (i.e., with the second inequality of (\ref{ineq1})) and the second inequality of (2) implies that $Q \not\in Y_0$. So, if there is one point $P\in Y_0$ closer to $P_0$ than point $A$, then no other point, but $A$, can precede $P$ in $Y_0$. Now, suppose that there is another point $Q\in Y_0$ that is also closer to $P_0$ than point $A$. This means that, as with $P$, no other point can precede $Q$ in $Y_0$. However, since both $P$ and $Q$ are in $Y_0$, one must precede the other, which is a contradiction. Therefore, in this case, the only point possibly closer to $P_0$ than point $A$ is the second element of $Y_0$.
\end{enumerate}
\end{description}

From all the previous three cases we may conclude that the closest point to $P_0$, in $Y_0$, is either the first or the second element of $Y_0$. This proves Lemma \ref{lemma1} for the Minkowski distance $d_1$.

To obtain proof for all other Minkowski distances $d_p$, $p>1$, we take into account the fact that the convex neighborhoods generated by the Minkowski distances possess a very straightforward order relation, where larger values of $p$ correspond to larger unit circles, as shown in Figure \ref{fig:Minkowski}. This ordering means that the sparsity effect within the $\Delta$ slab will be similar, but somewhat stronger, for larger values of $p$. Therefore, the precedent analysis of $d_1$ not only remains valid for $p>1$ but, in a sense, the corresponding geometric relations between elements of $Y_0$ are expected to be more ``tight'' for all other Minkowski distances. 
\qquad \end{proof}

\subsection*{Correctness}\label{subsec:correct}
To prove the correctness of the Basic-2.S4 algorithm we consider the following
\begin{Loop Invariant}
At the start of each iteration of the main loop, \emph{left} and \emph{right} are references to the points of \emph{$Y_L$} and \emph{$Y_R$},  respectively, with minimum y-coordinates, that still need to be compared with points on the opposite side. Also, $d_{min}$ corresponds to the value of the minimum distance found among all pairs of points previously checked.
\end{Loop Invariant}

We show that all three properties \textbf{Initialization}, \textbf{Maintenance}, and \textbf{Termination} hold for this loop invariant.

\begin{description}
\item[\textbf{Initialization}]At the start of the first iteration, the \textit{loop invariant} holds since \textit{left} and \textit{right} are references to the first elements of $Y_L$ and $Y_R$, and both arrays are y-sorted in ascending order (by construction). Also, no pair with points on opposite sides of line \textbf{\textit{l}} was checked yet, so the current minimum distance is the minimum between the left-side and right-side minimal distances $d_{Lmin}$ and $d_{Rmin}$, respectively. This value is stored in $d_{min}$.

\begin{figure*}[t]
\centering
\includegraphics[width=15cm]{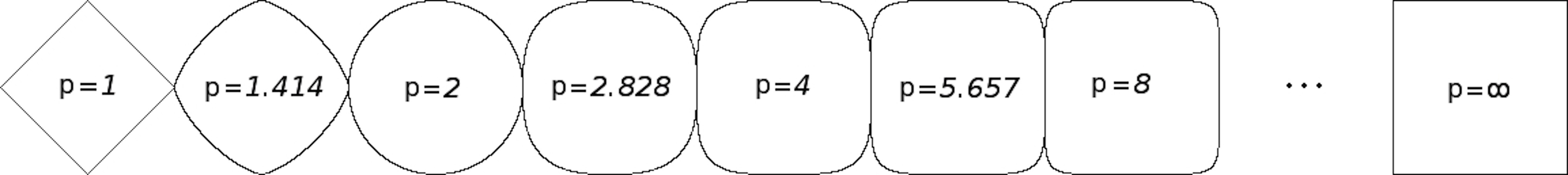}
\caption{Unit circles for various Minkowski p-distances.}
\label{fig:Minkowski}
\end{figure*}

\begin{figure*}
\centering
\includegraphics[width=15cm]{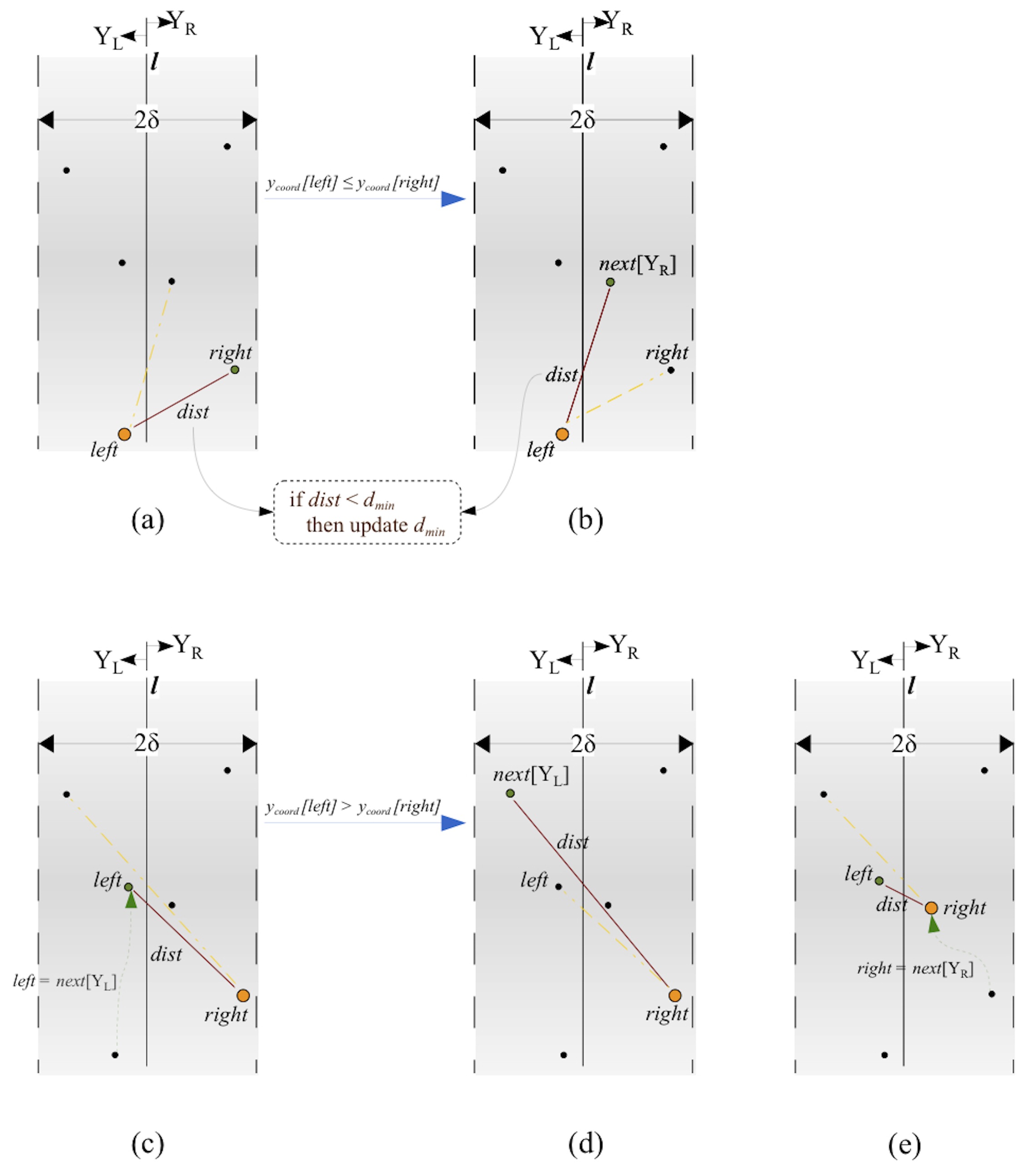}
\caption{Some iterations of the main loop in the Basic-2.S4 algorithm.
	\textbf{(a)} Compute distance between \textit{left} and \textit{right}, the first elements of $Y_L$ and $Y_R$, respectively.	
	\textbf{(b)-(c)} Point  \textit{left} is lower than point \textit{right} so, compute distance between  \textit{left} and the next element in $Y_R$. Update  \textit{left} to the next element in $Y_L$. Compute distance between  \textit{left} and  \textit{right}.	
	\textbf{(d)-(e)} Point  \textit{right} is lower than point  \textit{left} so, compute distance between  \textit{right} and the next element in $Y_L$. Update  \textit{right} to the next element in $Y_R$. Compute distance between  \textit{left} and  \textit{right}.}
\label{fig: iterations}
\end{figure*}

\item[\textbf{Maintenance}]Assuming that the \textit{loop invariant} holds for all previous iterations we now enter the next iteration. The first thing the loop does is computing the distance between the points referenced by \textit{left} and \textit{right} and, if that is the case, it updates the value of $d_{min}$. Next, the loop determines which of the two  points has smaller y-coordinate. Let us assume, without loss of generality, that in this iteration \textit{left} is the lowest point. Since $\textit{left} \in Y_L$, the algorithm checks to see if there is at least one more point in $Y_R$ (denoted by $next[Y_R]$), following the point $right$. If there is such point, the algorithm computes the distance between \textit{left} and $next[Y_R]$, and updates the value of $d_{min}$ accordingly.

By our hypothesis we know that \textit{right} and $next[Y_R]$ are the points on the right-hand side with the smallest y-coordinates that are still greater, at most equal, to \textit{left}'s y-coordinate. Therefore, and taking into account Lemma \ref{lemma1}, we conclude that the value of $d_{min}$ corresponds to the minimum between the previous minimal distance and the minimum distance computed for all pairs of points that contain the \textit{left} point.

The iteration ends by incrementing the reference  \textit{left} to the next point in $Y_L$, which means that the original left point will no longer be available for comparison. Note also that the  \textit{right} reference remains the same so that the corresponding point will still be compared with other points on future iterations (see Figure \ref{fig: iterations}). We may conclude that the new pair of references  \textit{left} and  \textit{right}, and the new value $d_{min}$ still satisfy the \textit{loop invariant}.

\item[\textbf{Termination}]The loop ends when one of the references,  \textit{left} or  \textit{right}, reaches the end of the corresponding array, $Y_L$ or $Y_R$, respectively. Let us assume, without loss of generality, that the \textit{left} reference reaches the end of the array $Y_L$ and terminates the loop. This means that it was the  \textit{left} reference that was incremented at the last iteration and so, it was this reference that corresponded to the lowest point. Accordingly, the loop computed the distances between  \textit{left} and the two closer, but not lower, points in $Y_R$ and updated $d_{min}$. As a consequence, all remaining pairs of points are composed by the point  \textit{left} and points that belong to the array $Y_R$ and lie in higher, more distant positions. Therefore, we have computed all distances between pairs of opposite points that may lie at a distance smaller than the current minimal distance, and so the value $d_{min}$ corresponds to the minimal distance between all pairs of points in $\Delta$. 
\end{description}
The Basic-2.S4 algorithm is correct and, therefore, the Basic-2 algorithm is also correct, for any Minkowski distance with $p\geqslant 1$.

\section{Empirical Time Analysis}\label{sec:empiric}

The Basic-2 algorithm has been implemented and tested with randomly generated inputs, starting with 125 thousand points and doubling the input size until 16 million points. For each input size, 50 independent executions were performed. For each specific generated input, we applied the Basic-2 algorithm as well as the standard divide-and-conquer algorithm, described in Cormen \textit{et al}~\cite{6}, which computes seven pairwise comparison for each point in the central slab. We refer to the standard algorithm as the Basic-7, following the naming convention presented in~\cite{7}. 

Both algorithms were implemented in the \textbf{C} programming language. 
The source code is available at {\color{blue}\url{http://w3.ualg.pt/~flobo/closest-pair/}}. 

\begin{figure*}
\centering
\includegraphics[width=15cm]{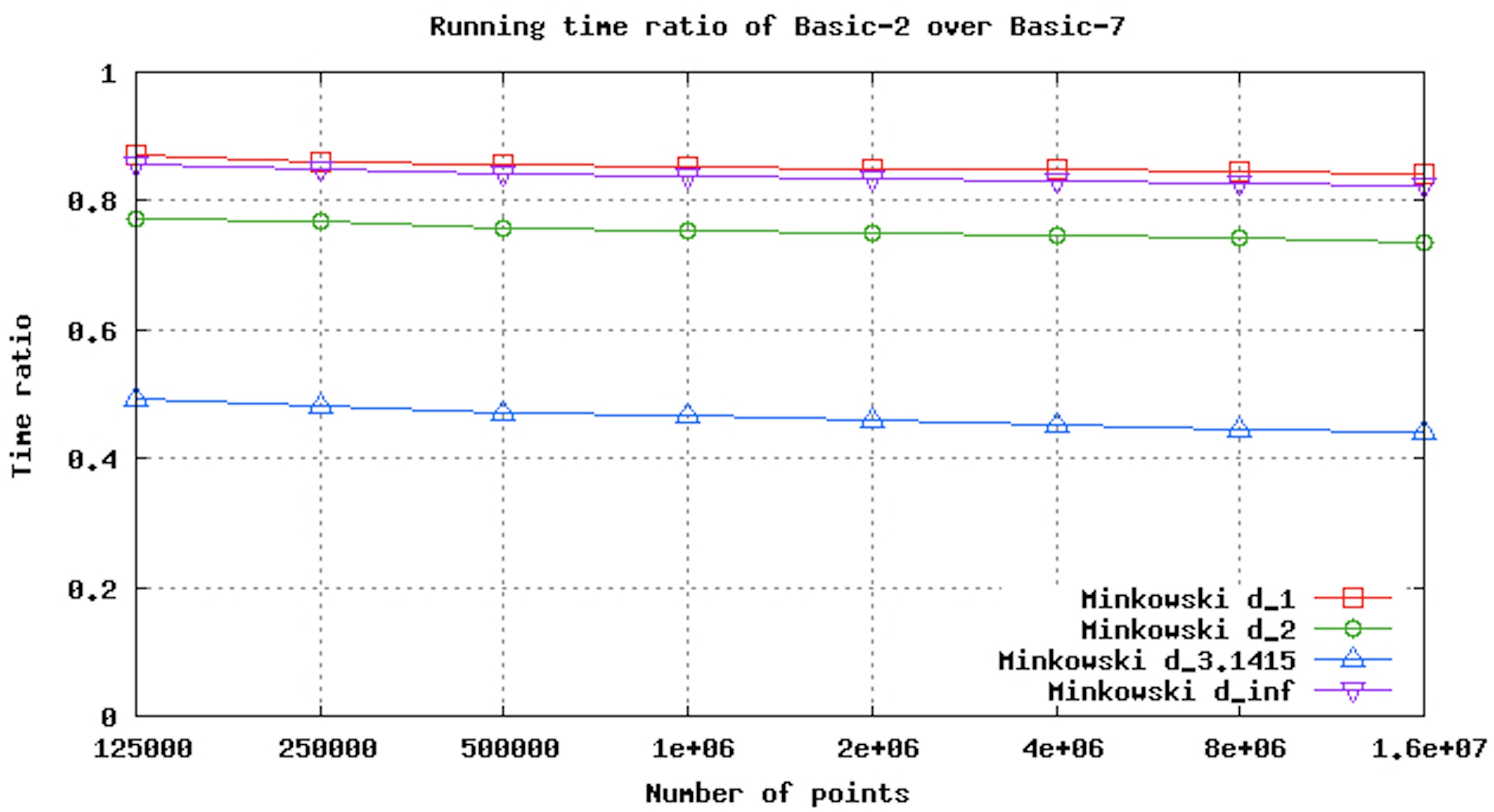}
\caption{Running time ratios of Basic-2 over Basic-7 for various Minkowski distances, averaged over 50 independent runs.}
\label{fig:timeratios}
\end{figure*}

The recursion stopped for a number of points less or equal to 10. The algorithms were tested with Minkowski distances $d_1$,  $d_2$, $d_{3.1415}$ and $d_\infty$. For each independent run, we measured the execution time of the entire algorithm. Figure~\ref{fig:timeratios} shows the ratio of the average execution time of Basic-2 over Basic-7, over the 50 independent runs. The coefficient of variation was less than 2.5\% for all Minkowski distances and input sizes tested, and for the larger input sizes it was always less than 0.07\%.

As expected, and in accordance with the results obtained by Jiang and Gillespie~\cite{7}, our simulation shows that  Basic-2 is faster than the standard divide-and-conquer algorithm, Basic-7. Although Basic-2.S4 introduces a few extra relational comparisons (see pseudocode in Figure~\ref{fig:pscode2}), these are negligible compared to the savings that occur due to the reduction in the total number of distance function calls.

The Basic-2 algorithm is about 20\% faster than the standard divide and conquer algorithm for the larger input sizes, when using the Minkowski distances $d_1$ and $d_\infty$. The speedup is more pronounced for the case when using the Minkowski distance $d_2$, with Basic-2 being nearly 36\% faster for the larger input sizes. The reason for the greater speedup when using Minkowski distance $d_2$ is because the computation of the distance function is more expensive in this case, and therefore, savings in distance function calls have a more profound effect on the overall execution time of the algorithm. This fact is fully confirmed by the results obtained when using the somewhat exotic distance $d_{3.1415}$. Due to the non-integer value of $p$, the cost of computing this kind of distance function is highly inflated and so it is possible to observe speedups over 100\% for the Basic-2 algorithm.

\section{Summary and Conclusions}\label{sec:final}
In this paper we analyzed the Basic-2 algorithm, which is an optimized version of the Bentley and Shamos procedure for the planar case where the computation of the distance $d_{LR}$ requires only two pairwise comparisons per point to find the closest pair within the central slab. The Basic-2 algorithm was first presented by Jiang and Gillespie~\cite{7}.

We show that, for this algorithm, only two pairwise comparisons are required in the combine step, for each point that lies in the central slab . This result and the subsequent correctness of the Basic-2 algorithm is shown for all Minkowski distances with $p\geqslant 1$.

We proved that the two comparisons per point are a minimum for the \textit{combine} step in the Basic-2 algorithm, for all Minkowski distances with $p\geqslant 1$. This result is a direct consequence of the strength of the sparsity condition, which is induced over the set of points in the plane by the knowledge of $d_L$ and $d_R$.

We consider that the generalization of this result to higher dimensions (in particular to the 3D space) is of interest, considering not only possible applications but also the theoretical significance of such an achievement. However, we note that, even in the 3D case this procedure may have rather less efficiency gains because of the lack of a natural order relation among the points lying in the central slab, and the consequent increase in the problem's complexity. 

%%%%%%%%%%%%%%%%%%%%%%%%%%%%% Main Text End %%%%%%%%%%%%%%%%%%%%%%%%%%%%%%%%%

\end{document}